\documentclass[11pt,amsmath,amssymb,nofootinbib]{revtex4} 
\usepackage{dcolumn}
\usepackage{bm}
\begin{document}
\title{Quantum fields, non-locality and quantum group symmetries}
\author{Michele Arzano}
\email{marzano@perimeterinstitute.ca}
\affiliation{Perimeter Institute for Theoretical Physics\\
31 Caroline St. N, N2L 2Y5, Waterloo ON, Canada}
\begin{abstract}
\begin{center}
{\bf Abstract}
\end{center}
We study the action of space-time symmetries on quantum fields in the presence of small departures from locality determined by dynamical gravity.  It is shown that, under such relaxation of locality the symmetries of the theory cannot be described within the usual framework of Lie algebras but rather in terms of non-co-commutative Hopf algebras or  ``quantum groups".  Similar ``quantizations" of space-time symmetries are expected to emerge in the low-energy limit of certain quantum gravity models and have been used to describe the symmetries of various non-commutative space-times.  Our result provides 
an intuitive characterization of the mechanism that could lead to the emergence of deformed co-products in models of quantum relativistic symmetries.  
\end{abstract}
\maketitle
\section{Introduction}
Symmetries play a prominent role in theoretical physics as they allow to establish constraints and make predictions for physical processes without knowing the detailed structure of the system under study.  Exact symmetries, however, are rarely realized in Nature.  Indeed it seems that our present knowledge of particle physics, up to the energies probed by experiments so far, owes much to a systematic, ``controlled", symmetry breaking. Moreover as we gain sensitivity in the probes which test the symmetry principles we assume as fundamental, often such exact symmetries appear only as approximations, at leading order in some physical scale, of more fundamental ones.  In some cases the need for such generalizations is suggested by a radical incompatibility between the framework in which the symmetries are described and certain fundamental aspects of the theory at hand.  A nice example of this is given by the transition from Galilean to Lorentz/Poincar\'e relativistic symmetries: the old Galilean framework for the description of symmetries was at odds with the intrinsic Lorentzian nature of Maxwell's theory of electro-magnetism.\\
Nowadays theoretical physics is facing a puzzle which might reflect a similar state of affairs.  Local quantum field theory (LQFT), even if extremely successful as an effective field theory in its range of validity, seem to grossly overcount the number of degrees of freedom in a given region of space.  In fact ``holographic" arguments predict a non-extensive scaling of the number of degrees of freedom for a given region of space determined by the area of the region \cite{Bousso:2002ju} while the degrees of freedom of local quantum fields scale with the volume. The emergence of non-locality is usually indicated as the cause for such tension.  Indeed, according to a common intuition (see \cite{Giddings:2005id, ArkaniHamed:2007ky} for recent discussions), locality (or microscopic causality) should be an approximate concept in quantum gravity since once the background metric becomes dynamical and is allowed to fluctuate the notion of space-like separation of two events potentially loses its meaning.  Our description of particle physics in terms of local field theory thus relies on the assumption that in an ideal setting even if an intrinsic non-locality is present its negligibly small effects will become important only in the ultraviolet where the effective description is supposed to break down anyway.  This expectation, however, turns out to be wrong \cite{ArkaniHamed:2007ky} when for example such tiny effects are amplified by a very large number of states.  
In these special cases the knowledge of how our effective theory is modified by non-locality becomes of vital importance.\\  In this Letter we argue that there is a qualitative difference between usual LQFT and quantum fields in the presence of an intrinsic non-locality.  In fact, while in the former case external space-time symmetries are described by the action of a Lie algebra on the asymptotic free states, in the presence of deviations from locality the characterization of such symmetries requires the use of non-trivial Hopf algebras known as ``quantum groups".\\  
In the next Section we will briefly recall how symmetries are described in the framework LQFT with particular emphasis on the relation between locality and the additive action of symmetry generators on asymptotic states.  In Section III we present our main argument, namely that the failure of strict locality requires a description of the symmetries of the theory in terms of non-cocommutative Hopf algebras (``quantum groups"), and we link our considerations to specific models of quantum group symmetries that have been studied in the literature.  The last Section contains a summary and outlook.
\section{Symmetries and local quantum fields}
Let us start by recalling the notion of locality (micro-causality) in quantum field theory and its implications for the symmetries of the theory.\\
Strictly speaking a field operator in LQFT, $\phi(x)$, is an ``operator valued distribution".  This means that the corresponding operator acting on the Hilbert space of the theory is obtained by smearing $\phi(x)$ with an appropriate $C^{\infty}$ test function 
\begin{equation}
\phi(f)=\int \phi(x) f(x)\, dx\, .
\end{equation}
If the function $f$ vanishes outside a bounded region $\phi(f)$ is a {\it localized} operator, if $f$ does not vanish but is fast decreasing with all its derivatives then $\phi(f)$ is a {\it quasi-local} operator \cite{Haag:1992hx}.  
A localized operator is said to be {\it local} iff
\begin{equation}
\label{loc2}
[\phi(f),\phi(g)]=0
\end{equation}
when the supports of the test functions $f$ and $g$ are space-like separated.  Now consider the translated operator
\begin{equation}
\phi(f;x)=U(x)\phi(f)U^{-1}(x)\, ,
\end{equation}
due to (\ref{loc2}) the commutator 
\begin{equation}
\label{comm4}
[\phi(f; t,\vec{x}_1),\phi(g; t ,\vec{x}_2)]
\end{equation}
vanishes for some finite value of $|\vec{x}_1-\vec{x}_2|$ if $\phi(f)$ and $\phi(g)$ are localized operators.  On the other hand  the commutator (\ref{comm4}) for quasi-local operators {\it does not vanish but falls off to zero faster than any inverse power of the spatial separation $|\vec{x}_1-\vec{x}_2|$} \cite{Haag:1992hx}.\\
The construction of the asymptotic states of a general LQFT relies exclusively on quasi-local operators. Indeed, in the Haag-Ruelle formalism \cite{Haag:1992hx, Landau:1970fs}, one constructs from appropriately smeared polynomials of the field operators a quasi-local operator $q(f,t)$ which creates a one-particle state $q(f,t)|\,0>=|f>$ with ``wavefunction" $<\vec{p}\,|f\,>=f(\vec{p})$ independent of $t$.
One can show that $q(f_1,t)...q(f_n,t)|\,0>$ has a strong limits for $t\rightarrow\pm\infty$
leading to the the asymptotic free states $|f_1,...,f_n>_{\mathrm{out}, \mathrm{in}}$  \cite{Haag:1992hx, Landau:1970fs}.  Under the assumption of {\it asymptotic completeness} the collections of $|f_1,...,f_n>_{\mathrm{out}, \mathrm{in}}$ span the entire Hilbert (Fock) space of physical states $\mathcal{F}(\mathcal{H})$ \cite{Haag:1992hx, Landau:1970fs}.  There will be a unitary operator, the $S$-matrix, such that
$|f_1,...,f_n>_\mathrm{out}=S\,|f_1,...,f_n>_\mathrm{in}$.  We are interested in the interplay between external (geometrical) symmetries and quantum fields.  
A key fact is that any symmetry describes certain properties which are preserved by the dynamics and thus {\it is fully characterized in terms of its action on the asymptotic, free state, configurations}.\\
Let us consider the simple example of a massive real scalar field.  A {\it symmetry transformation} of the theory is a one-parameter, continous, abelian unitary operator $U(\tau)$ in the space of physical states which commutes with the $S$-matrix and transforms one-particle states into themselves.  The symmetry transformation is said to possess a {\it generator} if it can be written as $U(\tau)=exp(iG\tau)$ with $G$ a self-adjoint operator. In LQFT such generators act on multiparticle states according to a generalized Leibnitz rule ({\it additive} action).  This last requirement is intimately related to the notion of locality.  To see this we look at how the symmetry generators are characterized in terms of the fundamental field observables.\\  
Given a local and locally conserved current $j_{\mu}(x)$ one can construct a symmetry generator corresponding to the ``formal charge" $Q$.  The latter can be defined as the limit 
\begin{equation}
Q=\lim_{T\rightarrow 0}\lim_{R\rightarrow \infty} j_0(f_R,f_T)
\end{equation}
of the ``partial charge" 
\begin{equation}
j_0(f_R,f_T)=\int dx f_R(\vec{x}) f_T(x_0) j_0(x)\,.
\end{equation}
with $f_R$ and $f_T$ appropriate smearing functions.  In particular $f_R(\vec{x})$ cuts the tails of the current for large spatial distances and $f_T(x_0)$ averages the current around the point $x_0=0$ (for details see \cite{Orzalesi:1970tx}).  
The question is whether or not the formal charge $Q$ defines a symmetry generator $G$.  The positive answer to this is given by a fundamental theorem due to Kastler, Robinson and Swieca (KRS) (see \cite{Orzalesi:1970tx} and references therein) which states that the commutator $[j_0(f_R,f_T),A]$ between the partial charge and any localized or quasi-local operator A is {\it independent of $f_R$ and $f_T$} for sufficiently large $R$.  In particular this is true for any quasi-local operator $A_{f_i}$ such that $A_{f_i}|\,0>=|\,f_i>$.  The KRS theorem allows one to define the action of the generator $G$ associated with the formal charge $Q$ through the {\it adjoint action}
\begin{equation}
\label{adjas}
GA |\,0>\equiv [Q,A]|\,0>=\lim_{T\rightarrow 0}\lim_{R\rightarrow \infty} [j_0(f_R,f_T),A]|\,0>\, ,
\end{equation}
as it guarantees that the limit in the last term exist and is independent of the particular choice of smearing functions.  One immediate consequence of the definition (\ref{adjas}) is that $G|\,0>=0$.  
Additivity of the action of $G$ immediately follows from the definition (\ref{adjas}) and the linearity of the commutator.  Such property is also manifest when one writes the generator in terms of the asymptotic creation and annihilation operators
\begin{equation}
G=\int d^3\vec{k}\,\eta(\vec{k})\,a^{\dagger}_{\mathrm{in,out}}
(\vec{k})\,a^{\phantom{\dagger}}_{\mathrm{in,out}}(\vec{k}) 
\end{equation}
where the kernels $\eta(\vec{k})$ characterize the action of the generator on one-particle states.  Indeed the expression above can be derived from the one-particle matrix elements of $G$
\begin{equation}
<\vec{k}|G|\vec{k}'>=\eta(\vec{k})\delta^{(3)}(\vec{k}-\vec{k}')
\end{equation}
and
\begin{align}
\label{adjo2}
[G,a^{\dagger}_{\mathrm{in,out}}(\vec{k})]&=\eta(\vec{k})a^{\dagger}_{\mathrm{in,out}}(\vec{k})\\
[G,a_{\mathrm{in,out}}(\vec{k})]&=-\eta(\vec{k}) a_{\mathrm{in,out}}(\vec{k})\,.
\end{align}
There is a nice algebraic way to characterize the additivity of a symmetry generator.  Let $G$ be an element of the Lie algebra $\mathfrak{g}$ describing the symmetries of the space on which our quantum fields live (in Minkowski space $\mathfrak{g}$ is simply the Poincar\'e algebra $\mathcal{P}$).  The one-particle Hilbert space $\mathcal{H}$ is an irreducible representation of $\mathfrak{g}$.  ``Multi-particle" (asymptotic) free states are given by appropriately symmetrized tensor products of $\mathcal{H}$.  What is the action of $G$ on such states or, in other words, how do we construct  representations of $\mathfrak{g}$ on tensor products of $\mathcal{H}$?  It turns out that the usual construction of tensor product representation for a Lie algebra $\mathfrak{g}$ is best understood in terms of the universal enveloping (UE) algebra $U(\mathfrak{g})$ associated to $\mathfrak{g}$.  In fact, UE algebras are an example of Hopf algebras which in turn are a generalization of standard (unital, associative) algebras.  Hopf algebras come equipped with additional structures which, among other things, allow one to properly define tensor product representations of $\mathfrak{g}$.  In particular the ``co-product" (or co-multiplication) $\Delta$ is a map $\Delta:U(\mathfrak{g})\rightarrow U(\mathfrak{g})\otimes U(\mathfrak{g})$ defined by
\begin{equation}
\label{copro}
\Delta(G)=G\otimes 1+1\otimes G
\end{equation}
where $1$ is the unit element of $U(\mathfrak{g})$.  Given two representations of $\mathfrak{g}$, $(\rho_1, \mathcal{H}_1)$ and $(\rho_2, \mathcal{H}_2)$ the tensor product representation $(\rho,  \mathcal{H}_1\otimes \mathcal{H}_2)$ is given by
\begin{equation}
\rho\equiv (\rho_1\otimes \rho_2) \Delta\, .
\end{equation}
The co-product (\ref{copro}) is just telling us that $G$ acts on a ``two-particle" state of $\mathcal{H}_1\otimes \mathcal{H}_2$ according to the Leibnitz rule i.e. the action of $G$ on such states is {\it additive}.  An important property of the co-product (\ref{copro}) is that it is {\it co-commutative} i.e.
\begin{displaymath}
\sigma\circ\Delta=\Delta\circ id
\end{displaymath} 
with $\sigma: U(\mathfrak{g})\otimes U(\mathfrak{g})\rightarrow U(\mathfrak{g})\otimes U(\mathfrak{g})$ the "flip" map $\sigma(a\otimes b)=b\otimes a$, $id$ the identity map and $\circ$ the composition of maps.  Hopf algebras possessing a co-commutative co-product are called {\it trivial}.  It is easy to see that co-commutative co-products lead to an additive action of $G$ on multi-particle states\footnote{The definition of an $n$-fold tensor product of representations of $\mathfrak{g}$ can be obtained by simply iterating the definition above.}.  
But that's not all.  The (trivial) Hopf algebra structure of the symmetries is present already at the one-particle level.  In fact, the action of $G\in U(\mathfrak{g})$ on the algebra of asymptotic creation and annihilation operators given by (\ref{adjo2}) is nothing but the ``adjoint action"
\begin{equation}
\label{adjact}
\mathrm{ad}_{G}(a_{\mathrm{in,out}})\equiv 
((id\otimes S)\Delta(G))\diamond a_{\mathrm{in,out}}
=[G,a_{\mathrm{in,out}}]
\end{equation}
where $S$ is the antipode map\footnote{Beside standard multiplication $m$,  unit map $\eta$ and the co-product $\Delta$ defined above a Hopf algebra possesses two additional maps, the co-unit $\varepsilon: U(\mathfrak{g})\rightarrow \mathbb{C}$ and the antipode $S: U(\mathfrak{g})\rightarrow U(\mathfrak{g}) $ satisfying the following axioms 
\begin{eqnarray*}
(\Delta\otimes id)\Delta&=(id\otimes \Delta)\Delta~~~~~~~\mathrm{co-associativity}\\
(id\otimes\varepsilon)\Delta&=(\varepsilon\otimes id)\Delta=id~~~~~~~\mathrm{co-unit}\\ 
m(S\otimes id)\Delta&=m(id\otimes S)\Delta=\eta\circ\varepsilon~~~~~~~\mathrm{antipode}\,.
\end{eqnarray*}} $S(G)=-G$ and $(F\otimes G)\diamond a=FaG$.  This shows how the Hopf algebra structure of the UE  $U(\mathfrak{g})$ associated to the Lie algebra of symmetries $\mathfrak{g}$ is hidden behind the familiar ``commutator" action of $G$ on linear operators on $\mathcal{F}(\mathcal{H})$.  It turns out that there exist ``quantum" deformations of UE algebras which lead to non-trivial Hopf algebras which are also known in the literature as {\it quantum groups}.  In the next section we will discuss how quantum deformations of UE algebras, in the context of quantum field theory, can be related to the presence of an irreducible non-locality.\\

\section{Quantum symmetries from quantum fields}
Consider the quantum theory of a massive real scalar field for which a set of asymptotic ``in" and ``out" states is given.  Under the assumption of asymptotic completeness these states span the full Hilbert (Fock) space of the theory $\mathcal{F}(\mathcal{H})$.  A unitary $S$-matrix connects the two sets of states.  From a ``purely" quantum mechanical point of view a symmetry of the theory is a mapping of rays of the Hilbert space which leaves invariant the transition probabilities.  According to Wigner's theorem (see e.g. \cite{Bargmann:1964zj}) space-time symmetries will be described by unitary operators $U$ on the asymptotic states.  Such operators commute with the S-matrix, map one particle states into themselves and leave the vacuum invariant.  
An infinitesimal transformation will be of the form $U=1+i\tau G$ with $\tau$ an infinitesimal parameter and $G$ the generator of the symmetry.  In particular, if we denote the action of the generator $G$ on an operator $A$ defined on $\mathcal{F}(\mathcal{H})$ with $G\vartriangleright A$, one has $<0\,|G\vartriangleright A|\,0>= 0$ \cite{Bargmann:1964zj}. The properties we described above are the minimal requirements that an external symmetry of our quantum fields has to fulfill.\\  We assume now that, according to the results of \cite{Giddings:2005id, Dittrich:2006ee, ArkaniHamed:2007ky}, the observables of the theory possess an intrinsic, irreducible, non-locality.  In \cite{Giddings:2005id} it is discussed how, starting from diffeomorphism invariant observables of an effective theory of quantum gravity, one could recover the familiar observables of local quantum field theory.  The conclusions reached in \cite{Giddings:2005id} seem to indicate a fundamental limitation in obtaining such local observables.  From a relational point of view in order to ``localize" an observable in a diffeo-invariant theory one needs a reference frame given by some dynamical field.  The question is whether or not one is able to define a reference frame which in a certain limit reproduces standard local observables of LQFT.  It turns out that to do so one has to pick a reference dynamical field which is itself intrinsically non-local \cite{Dittrich:2006ee}.  As discussed in \cite{ArkaniHamed:2007ky}, {\it dynamical} gravity is the crucial ingredient which changes the rules of the game.  The heuristic argument given in \cite{ArkaniHamed:2007ky} shows that switching on gravity has the effect of introducing an irreducible error in the measurement of quantum local observables which is {\it non-perturbative} in the coupling $G_N$ and is of the order $e^{-r^2/G_N}$ where r is the ``size" of the apparatus used in the measurement (or equivalently the spatial separation of two local observables).  The non-perturbative nature of the non-local effects discussed in \cite{ArkaniHamed:2007ky} suggests that, in a quantum gravitational setting, even though a sharp notion of locality is lost, weaker causality properties like those of quasi-local operators can be preserved.  Motivated by these considerations we assume that when fluctuations of the background space-time are present {\it the only sensible notion of locality in a theory of quantum fields is that of quasi-locality}.  As discussed in the previous Section this does not conflict with the construction and existence of asymptotic free states.  However the failure of ``strict" microscopic causality has deep consequences for the symmetries of the theory.  In fact local commutativity is a crucial ingredient in the proof of the KRS theorem (see Section 4.A of \cite{Orzalesi:1970tx}).  The presence of an irreducible non-locality renders void its statement i.e. $[j_0(f_R,f_T),A]$ and consequently $[Q,A]$ are not necessarily independent of $f_R$ and $f_T$ for large $R$.  Now, as we saw in the preceding Section, the action of a symmetry generator $G$ is characterized by its associated conserved charge $Q$.  The failure of the KRS theorem {\it does not} guarantee that (\ref{adjas}) consistently defines an operator $G$ associated to the charge $Q$ on the asymptotic states. Once the invariance of the vacuum is taken into account, a necessary condition for (\ref{adjas}) to be a consistent definition is that $<0\,|[Q,A]|\,0>=0$ for any quasi-local operator $A$.  If the generator of a given symmetry $G$ can not be defined in terms of  the ``adjoint" action $[Q,A]$ one then has 
\begin{equation}
\label{nladjo}
0=<0\,|G\vartriangleright A|\,0>\neq <0\,|[Q, A]|\,0>\, .
\end{equation} 
This is somewhat reminiscent of spontaneous symmetry breaking \cite{Goldstone:1962es} where one has a locally conserved current but for its associated charge $<0\,|[Q,A]|\,0>\neq 0$.  The crucial difference is that in our case we want to keep the {\it vacuum invariant} under the action of $G$.
Thus we see that the presence of an intrinsic non-locality, no matter how mild, requires a generalization of the ``adjoint" action $[Q,A]$.  Below we will show how {\it non-trivial Hopf algebras} naturally provide such a generaliztion.\\
Let us consider a charge $Q$ which fails to define an ``adjoint" action due to the intrinsic non-locality between the locally conserved current and any quasi-local operator.  Taking into account (\ref{nladjo}) and specializing to creation operators as in (\ref{adjo2}) we can write
\begin{equation}
\label{nladj}
<0\,|G\vartriangleright a^{\dagger}(\vec{k})|\,0>\equiv<0\,|[G, a^{\dagger}(\vec{k})]|\,0>+\alpha_1 E^{-1}_p F^{(1)}(\vec{k})+O(E^{-2}_p)
\end{equation}
where the non-local corrections are given by model-dependent functions of the momentum 
 $\vec{k}$ suppressed by inverse powers of the Planck energy $E_p$.  It turns out that {\it the ``deformed" adjoint action above can be effectively described by the ``semiclassical" expansion of the quantum adjoint action of a non-trivial Hopf algebra with deformation parameter $h=E^{-1}_p$}.  In particular the non-local behavior in (\ref{nladj}) is reproduced by a symmetry generators $G$ belonging to a non-cocommutative Hopf algebra obtained by a deformation of the universal enveloping (UE) algebra $U(\mathfrak{g})$ of a Lie algebra $\mathfrak{g}$.  These deformations are known as quantized universal enveloping (QUE) algebras and are one of the most important examples of quantum groups (see e.g  \cite{quantumgr,Tjin:1991me}).  
As mentioned at the end of last Section, QUE algebras exhibit non-trivial (non-cocommutative) co-products together with possible additional deformations of the co-algebra sector.  
The non-trivial co-product of a QUE algebra can be written in ``semiclassical" approximation \cite{Ruegg:1994bk} as
\begin{equation}
\Delta(G)=\Delta^{(0)}(G)+h\Delta^{(1)}(G)+O(h^2)
\end{equation}
with $\Delta^{(0)}(G)=G\otimes1+1\otimes G$, the trivial co-product.  Similarly for the deformed antipode one can write 
\begin{equation}
S(G)=S^{(0)}(G)+h\,S^{(1)}(G)+O(h^2)\, ,
\end{equation}
with $S^{(0)}(G)=-G$.
It is clear now that according to the definition of adjoint action given in (\ref{adjact}) the generator belonging to a QUE algebra will act through the ``quantum" adjoint action
\begin{align}
\mathrm{ad}_{G}(a^{\dagger}(\vec{k}))&=((id\otimes S)\Delta(G))\diamond a^{\dagger}(\vec{k})=\\
&[G,a^{\dagger}(\vec{k})]+h\left[((id\otimes S^{(1)})\Delta^{(0)}(G))\diamond a^{\dagger}(\vec{k})+((id\otimes S^{(0)})\Delta^{(1)}(G))\diamond a^{\dagger}(\vec{k})\right]+O(h^2)
\end{align}
which reproduces the ``symmetry breaking" of (\ref{nladj}) with the leading order terms of the deformed co-product and antipode determined by the model dependent, Planck-scale suppressed, non-local corrections.\\ 
QUE algebras have been studied extensively in recent years as candidate models for ``quantum" relativistic symmetries.  Two notable examples are the $\kappa$-deformed and $\theta$-``twisted" Poincar\'e algebras \cite{Lukierski:1992dt, Chaichian:2004yh}. Both ``quantum algebras" can be viewed as symmetries of different types of non-commutative space-times \cite{Chaichian:2004yh, Majid:1994cy, Agostini:2006nc}.  The $\kappa$-Poincar\'e algebra was originally obtained as a contraction of $U_q(\mathfrak{so}(3,2))$, the quantization of the UE algebra of the Anti-de Sitter algebra, with deformation parameter $q$.  In the contraction procedure the deformation parameter acquires dimension of a mass and is denoted by $\kappa$.  This type of deformation of the Poincar\'e algebra has gained popularity as a way to introduce a fundamental (planckian) length $\lambda=1/\kappa$ in a relativistic framework \cite{AmelinoCamelia:2000mn}.  In the last few years it has also been shown how such $\kappa$-symmetries naturally emerge in the description of the low-energy limit of certain $2+1$-dimensional quantum gravity models \cite{Amelino-Camelia:2003xp, Freidel:2005me}.  The $\kappa$-Poincar\'e algebra in its most studied version, the so-called ``biscrossproduct basis" \cite{Majid:1994cy}, exhibits both deformed co-product and antipode in the boost and translation sector (rotations are left untouched)
\begin{eqnarray}
\Delta(P_0)&=&P_0\otimes 1+1\otimes P_0\,\,\,\,\,\,\Delta(P_j)=P_j\otimes 1+e^{-P_0/\kappa}\otimes P_j \nonumber \\
\Delta(N_j)&=&N_j\otimes 1+e^{-P_0/\kappa}\otimes N_j+\frac{\epsilon_{jkl}}{\kappa}P_k\otimes N_l\, . \label{coprod}
\end{eqnarray}
and
\begin{eqnarray}
S(P_l)&=&-e^{\frac{P_0}{\kappa}}P_l\,\,\,\,\,\,\,S(P_0)=-P_0  \nonumber\\
S(N_l)&=&-e^{\frac{P_0}{\kappa}}N_l+\frac{1}{\kappa}\epsilon_{ljk}e^{\frac{P_0}{\kappa}}P_j M_k  \label{anti}\, .
\end{eqnarray}
The $\theta$-Poincar\'e algebra was obtained by ``twisting" the co-product of the UE algebra of the Poincar\'e algebra \cite{Chaichian:2004yh}.  In this case only the co-product for the boost-rotation sector is deformed while the antipodes are the same as in the standard case
\begin{equation}
\Delta(M_{\mu\nu})=M_{\mu\nu}\otimes 1+1\otimes M_{\mu\nu}-\frac{1}{2}\theta^{\alpha\beta}[g_{\mu\alpha}(P_{\nu}\otimes P_{\beta}-P_{\beta}\otimes P_{\nu})-g_{\nu\alpha}(P_{\mu}\otimes P_{\beta}-P_{\beta}\otimes P_{\mu})]\, .
\end{equation}
In the limits $\kappa\rightarrow\infty$ and $\theta\rightarrow 0$ one recovers in both cases the trivial Hopf algebra structure of the UE algebra of the Poincar\'e algebra.
$\theta$ and $\kappa$-deformed quantum fields are currently the subject of active study (see e.g. \cite{Balachandran:2007vx, Arzano:2007ef}).  Such theories exhibit several non-trivial features, most importantly they seem to lead to interesting behaviors in their multi-particle sectors hinting for possible deviations from usual statistics.    
\section{Conclusions}
We have discussed how a description of space-time symmetries in terms of quantum groups could arise in quantum field theory when the notion of strict locality is blurred by the effects of dynamical (quantum) gravity.  This result provides a physical motivation for the emergence of ``non-cocommutative co-products" which characterize the non-trivial Hopf algebra structure of the symmetries of certain non-commutative space-times.  Our argument suggests that these frameworks should in principle provide a ``finer" resolution than standard effective field theory in describing processes in which the latter ceases to be a good approximation. 
An important task left for future studies is to investigate the non-local behaviors of different effective quantum gravity models and the relations with their counterparts in terms of space-time quantum group symmetries.

\begin{acknowledgments}
I am indebted to Bianca Dittrich for several stimulating conversations and for comments on a preliminary draft of this letter and to Giovanni Amelino-Camelia for a critical reading of the manuscript.  I would also like to thank Florian Koch, Tim Koslowski and Giuseppe Policastro for discussions and useful remarks.\\
Research at Perimeter Institute for Theoretical Physics is supported in
part by the Government of Canada through NSERC and by the Province of
Ontario through MRI.
\end{acknowledgments}

\end{document}